\documentclass[preprint2]{aastex}
\usepackage{amssymb}
\usepackage{ctable}
\usepackage{morefloats}
\usepackage{graphicx,epsfig,fancyhdr,psfig,rotating,amsmath,epsf,txfonts,natbib,epstopdf,multirow,appendix}
\usepackage[colorlinks, citecolor=blue]{hyperref}
\usepackage{longtable}
\usepackage{lineno}
\def \I {\uppercase\expandafter{\romannumeral1}}
\def \II {\uppercase\expandafter{\romannumeral2}}
\def \III {\uppercase\expandafter{\romannumeral3}}
\def \XIV {\uppercase\expandafter{\romannumeral14}}
\def \arcsec{$^{''}$}

\def \bps {BPs\ }

\graphicspath{{Figure/}}

\begin{document}
\title{Magnetic flux supplement to coronal bright points}
\author{{\sc Chaozhou Mou\altaffilmark{1},
Zhenghua Huang\altaffilmark{1}, Lidong Xia\altaffilmark{1}, Maria S. Madjarska\altaffilmark{2}, Bo Li\altaffilmark{1}, Hui Fu\altaffilmark{1}, Fangran Jiao\altaffilmark{1}, Zhenyong Hou\altaffilmark{1}}}
\altaffiltext{1}{Shandong Provincial Key Laboratory of Optical Astronomy and Solar-Terrestrial Environment, Institute of Space Sciences, Shandong University, Weihai, 264209 Shandong, China\quad  z.huang@sdu.edu.cn}
\altaffiltext{2}{Armagh Observatory, College Hill, Armagh BT61 9DG, UK}

 \date{Received date, accepted date}

\begin{abstract}
 Coronal bright points (BPs) are associated with magnetic bipolar features (MBFs) and magnetic
   cancellation. Here, we investigate how BP-associated MBFs form and how the consequent
magnetic cancellation occurs. We analyse longitudinal magnetograms from the Helioseismic and
Magnetic Imager to investigate the photospheric magnetic flux evolution of 70 BPs.  From images
taken in the 193\,\AA\ passband of  the Atmospheric Imaging Assembly (AIA) we dermine that  the
BPs' lifetimes vary from 2.7 to 58.8 hours.  The formation of the BP MBFs is found to involve
three processes, namely emergence, convergence and local coalescence of the magnetic fluxes.
The formation of a MBF can involve more than one  of these processes. Out of the 70 cases, flux
emergence is the main process of a MBF buildup of 52 BPs, mainly convergence is seen in 28, and
14 cases are associated with local coalescence. For MBFs  formed by bipolar emergence, the time
difference between the flux emergence and the BP appearance in the AIA 193\,\AA\ passband
varies from 0.1 to 3.2 hours with an average of 1.3 hours. While magnetic cancellation is found
in all 70 BPs, it  can occur in three different ways: (\I) between a MBF and small weak
magnetic features (in 33 BPs); (\II) within a MBF with  the two polarities moving towards each
other from a large distance (34 BPs); (\III) within a MBF whose two main polarities emerge in
the same place simultaneously (3 BPs). While a MBF builds up the skeleton of a BP, we find that
the magnetic activities responsible for the BP heating may involve small weak fields.
\end{abstract}


\maketitle

\section{Introduction}
\label{sect_intro} Coronal bright points (BPs) are small (on average 20\arcsec--30\arcsec) and
short-lived (from a few minutes to a few tens of hours) bright structures, ubiquitously found in
the solar corona. They are believed to be the signature of a direct energy deposition in the
upper solar atmosphere \,\citep{1993SoPh..144...15W,2007ApJ...670.1401M}. Coronal BPs were first
identified in X-ray images and were named X-ray bright points (XBPs) in the
1970s\,\citep{vaiana1970a}. They have an average lifetime of 8 hours in X-rays
\citep{1974BAAS....6..265T,1976IAUS...71..145G,1976SoPh...49...79G}.  When observed in
Extreme Ultra-violet (EUV)\,\citep[e.g.][etc.]{1981SoPh...69...77H,1988BAAS...20Q.977H,2001SoPh..198..347Z,madj2012},
they are
 found to have a lifetime that
varies from a few minutes to a few days with an average of 20 hours\,\citep{2001SoPh..198..347Z}.
\bps consist of multiple small-scale (a few arcseconds) and rapidly-evolving (a few minutes)
loops\,\citep[e.g.][]{1979SoPh...63..119S,1990ApJ...352..333H,2004ESASP.575..535U}.

\par
Many
studies\,\citep[e.g.][etc.]{1971IAUS...43..397K,1975BAAS....7Q.350G,1975IAUS...68...23G,1976IAUS...71..145G,1977SoPh...53..111G,
1993SoPh..144...15W,
2001SoPh..201..305B,2003A&A...398..775M,2012A&A...548A..62H,2013SoPh..286..125C} have revealed
that BPs are associated with magnetic bipolar features (MBFs). A MBF is a pair of
opposite-polarity magnetic features observed in magnetogram data. The relationship between
magnetic flux emergence and BPs has been studied in the past  but only limited case studies were
reported during the Solar Heliospheric Observatory (SoHO) era. \citet{1977SoPh...53..111G}
compared BPs observed in X-rays with ephemeral active regions (i.e. emerging bipolar regions) and
found a weak correlation. \citet{1979SoPh...64...93M} argued that the discrepancy can be
explained if the BPs are associated with short-lived ephemeral active regions that might not have
been detected in the low temporal resolution data. BPs are also strongly associated  with
magnetic
cancellation\,\citep[e.g.][etc.]{1993SoPh..144...15W,2001SoPh..201..305B,2003A&A...398..775M,2012A&A...548A..62H}.
A study by \citet{1993SoPh..144...15W} found that 18 of 25 BPs were associated with magnetic
cancellation. They also found that BPs have a stronger connection with magnetic cancellation
rather than flux emergence. From high resolution Solar Optical Telescope (SOT) magnetograms,
 \citet{2012A&A...548A..62H} detemined that all 28 BPs  were associated with flux
emergence followed by magnetic field cancellation.

\par
The strong connection between BPs and magnetic field cancellation has been considered as evidence
of magnetic reconnection occurring in  BPs. \citet{1999ApJ...510L..73P} suggested that all the
energy losses of a BP are in fact replenished by magnetic energy. Magnetic reconnection in BPs
has also been supported  by  a magnetic field dynamic reconfiguration of
BP\,\citep{2008A&A...492..575P, 2011A&A...526A.134A, 2012ApJ...746...19Z}. Recently,
\citet{2014A&A...568A..30Z} suggested that interchange reconnection might occur between two close
chambers of a BP.  \citet{1994ApJ...427..459P} proposed a scenario where the reconnection results
from the converging motion of magnetic polarities. According to this model, the interaction
distance has to be less than a certain value in order to trigger the appearance of a BP. This
model  was also further developed  by \citet{1994SoPh..153..217P} and
\citet{2006MNRAS.366..125V,2006MNRAS.369...43V}, and it was linked to various observations
\citep[e.g.][etc.]{2001SoPh..201..305B,2003A&A...398..775M,2012ApJ...746...19Z,2012A&A...548A..62H}.

\par
Although BPs have been studied since the 1970s, many open questions remain. Firstly, how does a
BP-associated MBF form? This question has not been answered because high-resolution and high-cadence magnetic field observations in the past were taken for a limit period of time (hours) and had a limited FOV. For instance, the high-resolution observations by the Magnetic Doppler Imager onboard SoHO were restricted only  to the  disk center region, had a limited field-of-view (FOV) and were only occasionally coordinated with the highest spatial resolution (in the past) coronal imager, the Transition Region And Coronal Explorer (TRACE). In contrast, the Helioseismic and Magnetic Imager (HMI) onboard  the Solar Dynamics Observatory (SDO) provides constantly-taken full-disk longitudinal magnetograms at 45\,s cadence together with a 12\,s cadence imaging of the solar atmosphere by the Atmospheric Imaging Assembly (AIA). Secondly,  it is not yet known whether magnetic cancellation associated with BPs  occurs
only between the two main polarities of  the BP MBF  or whether small-scale magnetic fluxes that
emerge/evolve in the vicinity of the MBF are also involved? These questions are crucial for the
understanding on  how BPs are formed and heated to coronal temperatures.  In the present study,
we use the AIA data to identify BPs and  analyse
 HMI magnetograms to investigate their magnetic field evolution.  The article is organised as follows: In
Section \ref{sect_obser}, we describe the observations and the data analysis. The results are
presented in Section \ref{sect_resu}. The discussion and conclusions are given in
Section\,\ref{sect_conc}.

\section{Observations and analysis}
\label{sect_obser} The data used in the present study were taken by the AIA
\citep{2012SoPh..275...17L} and the HMI\,\citep{2012SoPh..275..207S} onboard
SDO\,\citep{sdo2010}. The AIA data have   a 12\,s cadence and 1.2\arcsec spatial resolution. The
HMI longitudinal magnetograms used in this study are  obtained at  45\,s cadence and have
$\sim$1\arcsec spatial resolution. The noise level of the HMI data is $\sim$10\,Mx\,cm${^{-2}}$
\citep{2012SoPh..279..295L}. The data used in this study were taken from 2010 December 31,
12:00\,UT to 2011 January 04, 00:00 UT, i.e. 84 hours in total.

\par
The AIA and HMI data were reduced using the IDL routine \textit{aia\_prep.pro}. The AIA
1600\,\AA\ and 304\,\AA\ images were used to examine the alignment between the AIA 193\,\AA\
images and the HMI magnetograms. We found that \textit{aia\_prep.pro} has performed an alignment
of the two instruments with an accuracy better than 3\arcsec.  We first identified each BP in the
AIA 193\,\AA\ images, then searched for its associated magnetic features within 20\arcsec\ around
the BP. We followed the evolution of each BP in the imaging data and the magnetograms  to
determine which magnetic features are the ones that are responsible for the BP formation and
evolution.

\par
We aim to study the magnetic field evolution of the BPs from their birth until their  full
disappearance. We visually  identified  70 BPs. These BPs are well representative as their total
emission and their associated magnetic flux distribute over  a large range of  values (see
Section\,\ref{sect_general} for detail). To follow the evolution of  the magnetic features, we
first used  automatic techniques, e.g. the Southwest Automatic Magnetic Interpretation Suite
\citep[SWAMIS,][]{2007ApJ...666..576D,2008ApJ...674..520L,2010ApJ...720.1405L,2013ApJ...774..127L}.
SWAMIS is powerful in tracking features that satisfy a set of conditions, e.g. minimum and maximum of flux density, size and lifetime, etc. However, many cases in the present study involve small-scale magnetic
elements that have various sizes and flux densities that can not be recognized by a set of input
parameters in the automatic algorithms. Thus, the visual analysis was best suited for the present work.

\section{Results and discussion}
\label{sect_resu} In all 70 cases,  a BP detected in the AIA 193\,\AA\ images is associated with
a MBF.  Usually, MBFs  appear  as two clusters of opposite-polarity magnetic features in the HMI
magnetograms and are found  in the vicinity of the BPs. Please note that identification of a MBF
in our study is based on the presence of a BP,  i.e. a BP is firstly identified and the
corresponding MBF is identified after. For each case we traced back the evolution of the MBF and
the surrounding flux concentrations  12 hours prior to the birth of the BP in order  to study how
a MBF is formed, and how it evolves and interacts with the surrounding magnetic features.

\subsection{General characteristics of BPs and their associated MBFs}
\label{sect_general}
\begin{figure*}[!ht]
\centering
\includegraphics[clip,width=8.5cm]{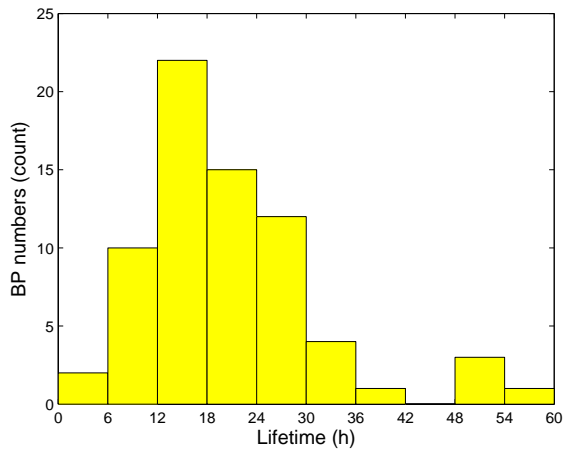}
\caption{Statistical histogram of the lifetime of 70 EUV BPs. The bin size is 6 hours.}
\label{fig_lifetime}
\end{figure*}

Once a BP was visually identified, its AIA 193\,\AA\ lightcurve was produced. A BP lightcurve is normally spiky but the long-term evolution clearly presents a fast rising phase and a fast decaying phase \citep{2001SoPh..198..347Z}. Based on this, the birth and death of a BP were determined from the start of the rising phase (i.e. a sharp intensity increase with respect to the background emission) and the end of the decaying phase (i.e. a drop of the emission to the background level), respectively. The lifetimes of the 70  BPs are  listed in Table\,\ref{tab_analysis}, and a histogram is shown in
Fig.\,\ref{fig_lifetime}. In AIA 193\,\AA, the BP  lifetime ranges from 2.7 to 58.8 hours, with
an average of 20.9 hours. This is consistent with the results of\,\citet{2001SoPh..198..347Z}, who
found a 20 hour average lifetime based on  a statistical analysis of 48 EUV BPs observed in the
EIT 195\,\AA\ passband that has  a similar temperature response to the AIA 193\,\AA\ channel.

\par
\begin{figure*}[!ht]
\centering
\includegraphics[clip,width=8.5cm]{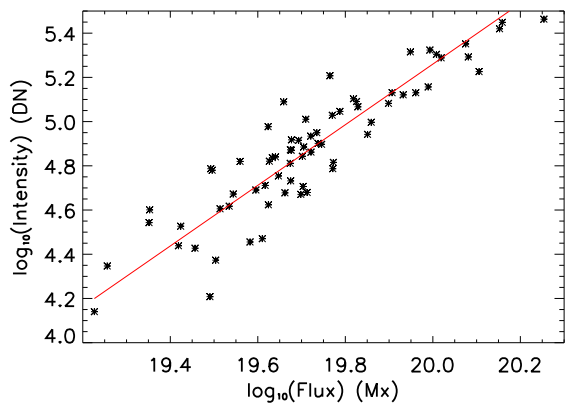}
\caption{Distribution of the total AIA 193\,\AA\ radiance of the 70 BPs vs. their total unsigned
magnetic flux. Both the AIA 193\,\AA\ radiance and the magnetic flux are measured at the time when
the BPs reach their radiation peak. The red solid line is the linear fit of the data points.}
\label{fig_peak}
\end{figure*}

Similarly to \citet{2003ApJ...598.1387P}, we investigate the relationship between the emission of the BPs and their total unsigned magnetic flux. In Fig.\,\ref{fig_peak}, we display a plot of the total AIA 193\,\AA\ radiances of the 70 BPs vs. the total unsigned magnetic fluxes of their associated MBFs. The measurements were taken at the peak of the BP
radiation. The linear fit given in Fig.\,\ref{fig_peak}
indicates a power-law approximation of $L_x\propto\Phi^{1.37}$, where $L_x$ is the radiance of a BP and $\Phi$ is its total unsigned magnetic flux of the associated MBF.

\par
\begin{figure*}[!ht]
\centering
\includegraphics[clip,width=17cm]{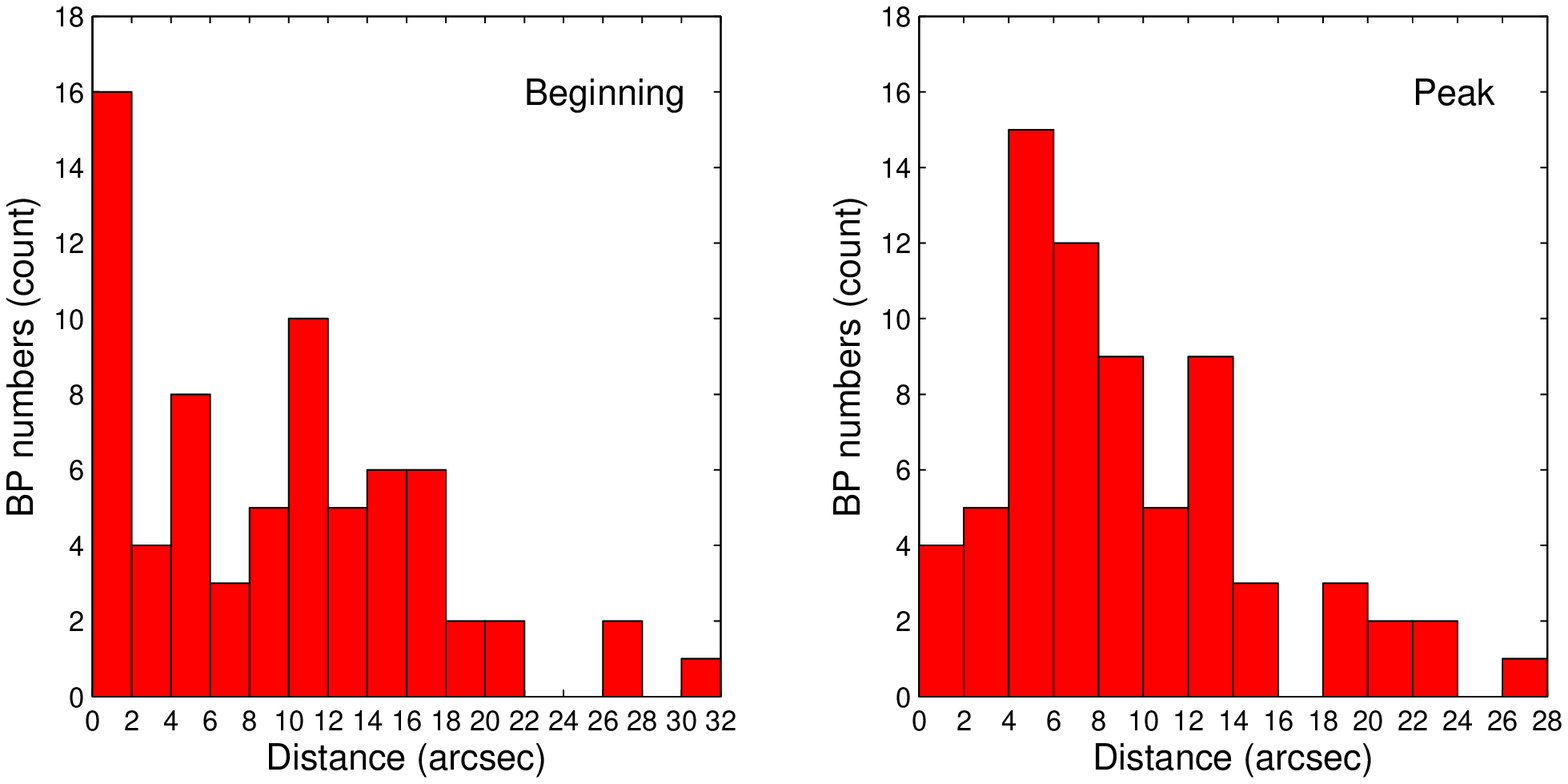}
\caption{Statistical histogram of the distances of the two main polarities of the MBFs of the 70
BPs at their first appearance (left) and when they are at their radiation peak (right).}
\label{fig_dis}
\end{figure*}

As discussed in Section\,\ref{sect_intro}, the distance between the two main polarities of a MBF
can be used to  test BP models, e.g. the convergence model by \,\citep{1994ApJ...427..459P}. To
obtain objective measurements of the distance between  the two polarities of the MBFs,  we
determined the separation  between the closest points of  the 30\,Mx\,cm$^{-2}$ contours of the
two opposite polarity magnetic features. If the two contours were  in contact, the distance was
defined as zero. In Fig.\,\ref{fig_dis}, we show the statistical results for all 70 BPs at their
first appearance (left panel) and at their radiation peak (right panel) in AIA 193\,\AA.
Considering the time when the BPs first appear, the distance is  in the range from  0 to
31.5\arcsec. A zero distance is found in 16 BP-associated MBFs, 15 of which are related  to
bipolar flux emergence.
 While the BPs are at
their radiation peak, the distance spreads from 0\arcsec to 26.8\arcsec with a distribution peak
at $\sim$ 5\arcsec. The distance of the two main polarities of the BPs will be further discussed
in the following sections regarding the formation and evolution of MBFs.

\subsection{On the formation of MBFs associated with BPs}
\label{sect_mbf_form}
 In the present
work, a MBF was visually traced back 12 hours prior to the first appearance of a BP in
AIA~193~\AA.  We found that the BP-associated MBFs can be formed via three ways: emergence,
convergence and local coalescence.  Magnetic flux emergence is a phenomenon which represents the
appearance of a magnetic feature in  magnetogram data. Convergence is a process where two
pre-existing opposite magnetic polarities that are initially found at far away distances move
towards each other, and a  BP would appear in the AIA 193\,\AA\ images once the polarities move
closer than a certain distance. It is possible that the two opposite polarities of a MBF are
already connected while the distance between the polarities is still quite large. In this study
we found that such a MBF can not produce a BP in the AIA 193\,\AA\ if convergence is not at work.
To clarify this issue, we believe that one would need a reliable field extrapolation that is
worthy to be tackled in future studies. Local magnetic coalescence is a phenomenon where a
magnetic feature is formed by the merging of  small-scale magnetic features (with a size close to
the instrumental resolution limit) of the same sign. However, our analysis shows that  the  two
polarities of a MBF are not usually formed in a single way. For example, in some cases, a
polarity formed by emergence or local coalescence moves towards a pre-existing one to form a MBF
that is associated with a BP.

\par
\begin{figure*}[!ht]
\centering
\includegraphics[trim=0.5cm 0.5cm 0.5cm 0.5cm,clip,width=8.5cm]{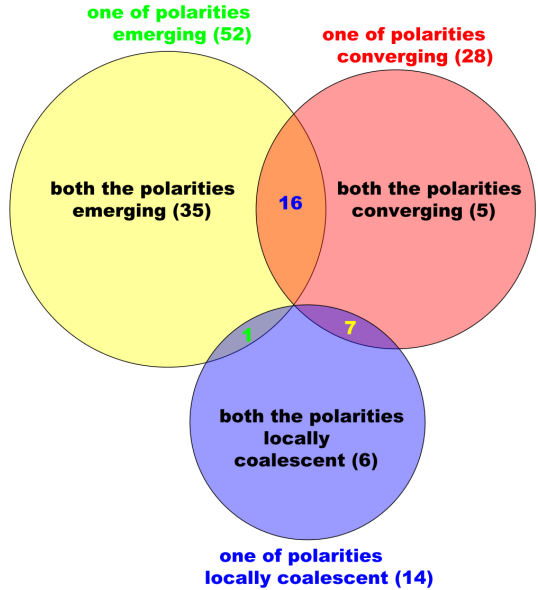}
\caption{Statistical Venn diagram of  the formation of the MBFs associated with the 70 BPs.}
\label{fig_venn_formation}
\end{figure*}

In Table\,\ref{tab_analysis}, we list the formation methods for the MBFs of all 70 BPs. A Venn
diagram of the statistical results is given in Fig.\,\ref{fig_venn_formation}. We found that in
46 cases the two polarities are formed in the same way, in which 35 are emergence, 5 are
convergence and 6 are local coalescence. The MBF formation for the remaining 24 cases involved
more than one process.  In summary, out of 70 BPs, emergence is found in 52 cases, convergence in
28 cases and local coalescence  in 14 cases. In the following  sections we give examples of these
three formation processes.

\begin{figure*}[!ht]
\centering
\includegraphics[clip,trim=0cm .5cm 0cm 0cm,width=17cm]{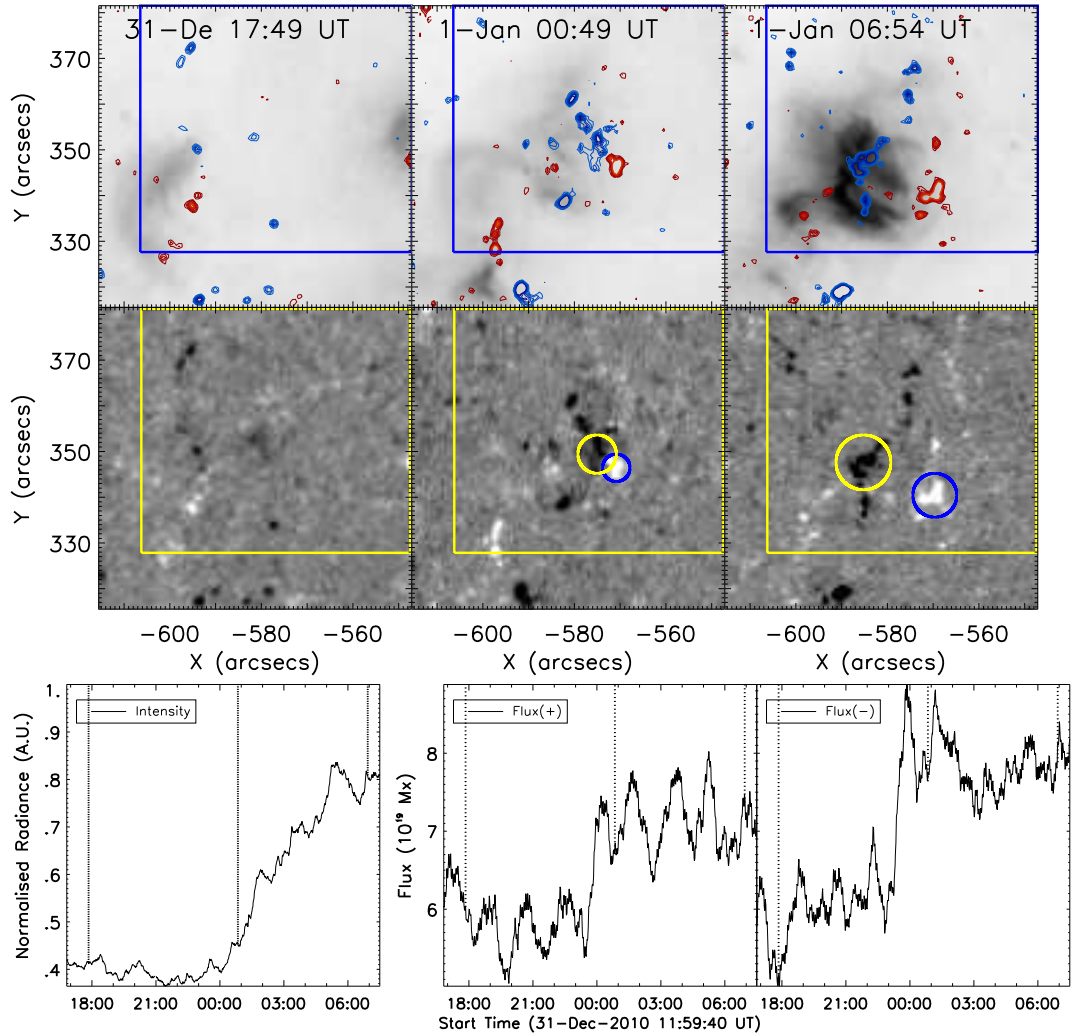}
\caption{The formation process of a MBF and the associated BP (No.~1 in Table\,\ref{tab_analysis}),
which is formed by bipolar emergence. The top row gives the BP region as seen in AIA 193\,\AA\ at
the indicated time. The middle row shows the corresponding HMI magnetograms. The bottom row shows the normalised
AIA 193\,\AA\ lightcurve (left), and the positive (middle) and negative (right) magnetic fluxes obtained
in the blue and yellow boxes marked in the top two rows. The contours of positive (red) and negative
(blue) magnetic flux density are over-plotted in the top row. The yellow and blue circles over-plotted on the
magnetograms denote the locations of the negative and positive polarities of the MBF. The dotted lines
over-plotted in the bottom row denote the time of the images shown on the top and middle rows. An
animation is enclosed online.}
\label{fig_emerg}
\end{figure*}

\begin{figure*}[!ht]
\centering
\includegraphics[trim=0.5cm 0cm 0.5cm 0.5cm,clip,width=8.5cm]{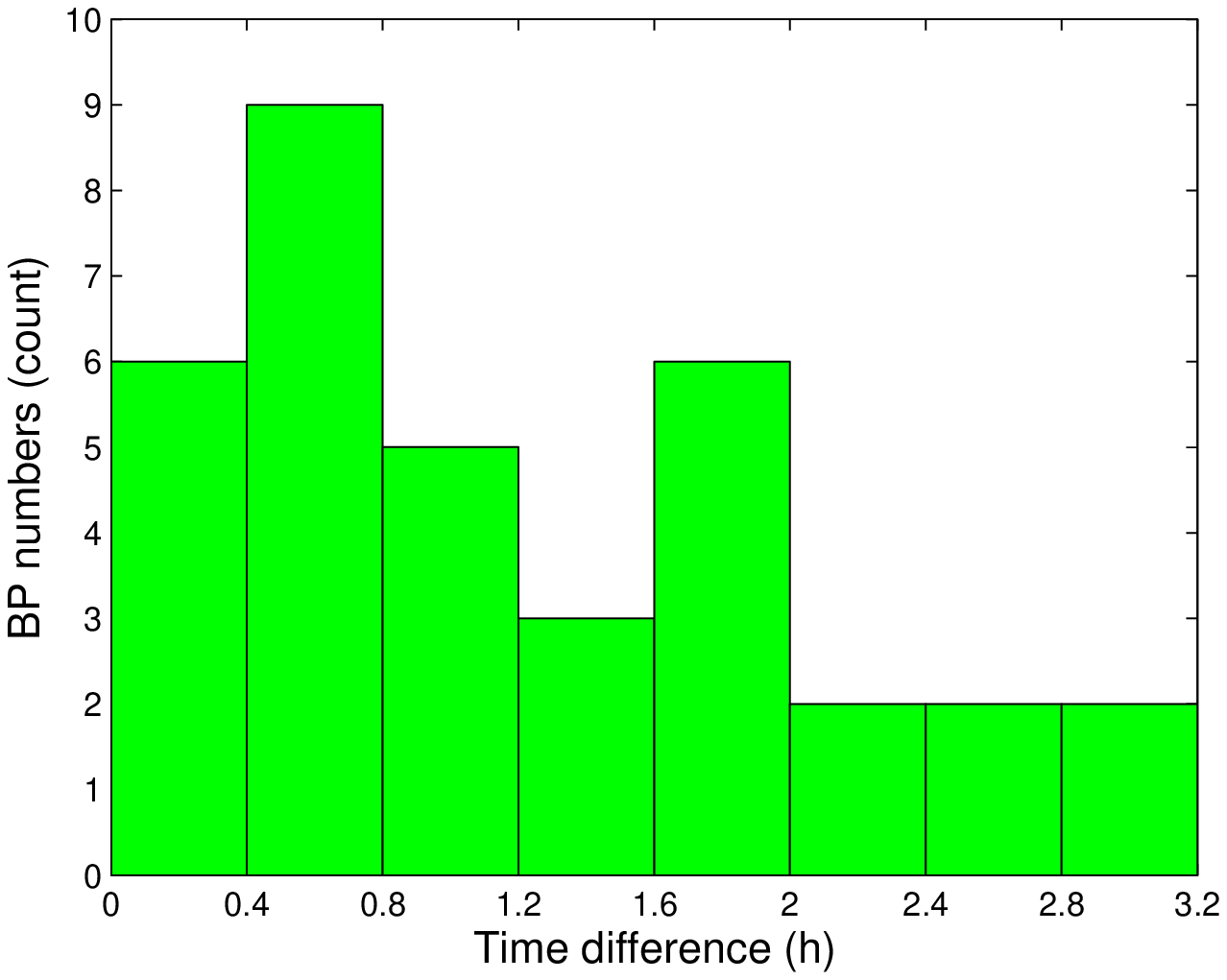}
\caption{Statistical histogram of the  time difference between the emergence of MBFs and the appearance of BPs in the cases when the MBF is formed by one bipolar emergence.}
\label{fig_tdiff}
\end{figure*}

\subsubsection{A MBF formed by bipolar emergence}
Fig.\,\ref{fig_emerg} presents the  formation process of a MBF associated with the BP listed as
No.~1 in Table\,\ref{tab_analysis}. The BP is first  detected at 00:04\,UT on January 1 and
disappears at 13:19\,UT on January 2, with a total duration of 37.3 hours. The BP MBF is formed
through emergence, i.e. the two opposite polarities emerge at the same location and then separate
from each other. The MBF emergence begins at 21:34\,UT on  2010 December 31, about two and
half hours prior to the first appearance of the BP in AIA 193\,\AA. The variations of both, the
negative and the positive fluxes,  show a rapid increase around this time. At the beginning, the
two polarities are attached to each other (see the snapshot at 00:49\,UT) while a fuzzy structure
of the BP is already seen in AIA 193\,\AA. The polarities start moving away from each other while
the magnetic features are becoming larger.  At the same
 time the BP appears in AIA 193\,\AA\  (see the top row of
Fig.\,\ref{fig_emerg}). An animation attached to Fig.\,\ref{fig_emerg} clearly demonstrates this
process. The AIA 193\,\AA\ lightcurve of the BP shows that a sharp intensity increase starts at
approximately 00:00\,UT and reaches its peak at 06:54\,UT on January 1 (see left panel of the
bottom row of Fig.\,\ref{fig_emerg}). At 06:54\,UT the BP is at its peak intensity and the
distance between the two main  polarities of the MBF is $\sim$11\arcsec.

\subsubsection{Time difference between  emergence of MBFs and appearance of BPs}
From 70 BPs, 35 are associated with MBFs formed by bipolar emergence process. Here, we measured
the time difference between the flux emergence and the BP appearance in these 35 cases.
A MBF is considered as emerging when it appears and has a flux density above 30\,Mx\,cm$^{-2}$.
A statistical histogram of the time difference is shown in Fig.\,\ref{fig_tdiff}. We found that the time difference varies from 0.1 to 3.2 hours with an average of 1.3 hours. The reasons for this time difference
might be twofold. First, it takes a certain amount of time for heating the BP plasma to a
temperature level at which  the AIA 193\,\AA\ channel is sensitive. Second, it is possible that
it takes some time after  the flux emergence that the physical process heating the plasma (thus
producing a BP) starts to operate. Since the time scale of plasma heating is normally very small, the second
reason is the most plausible, though the first one might work in some cases.

\begin{figure*}[!ht]
\centering
\includegraphics[clip,trim=0cm .5cm 0cm 0cm,width=17cm]{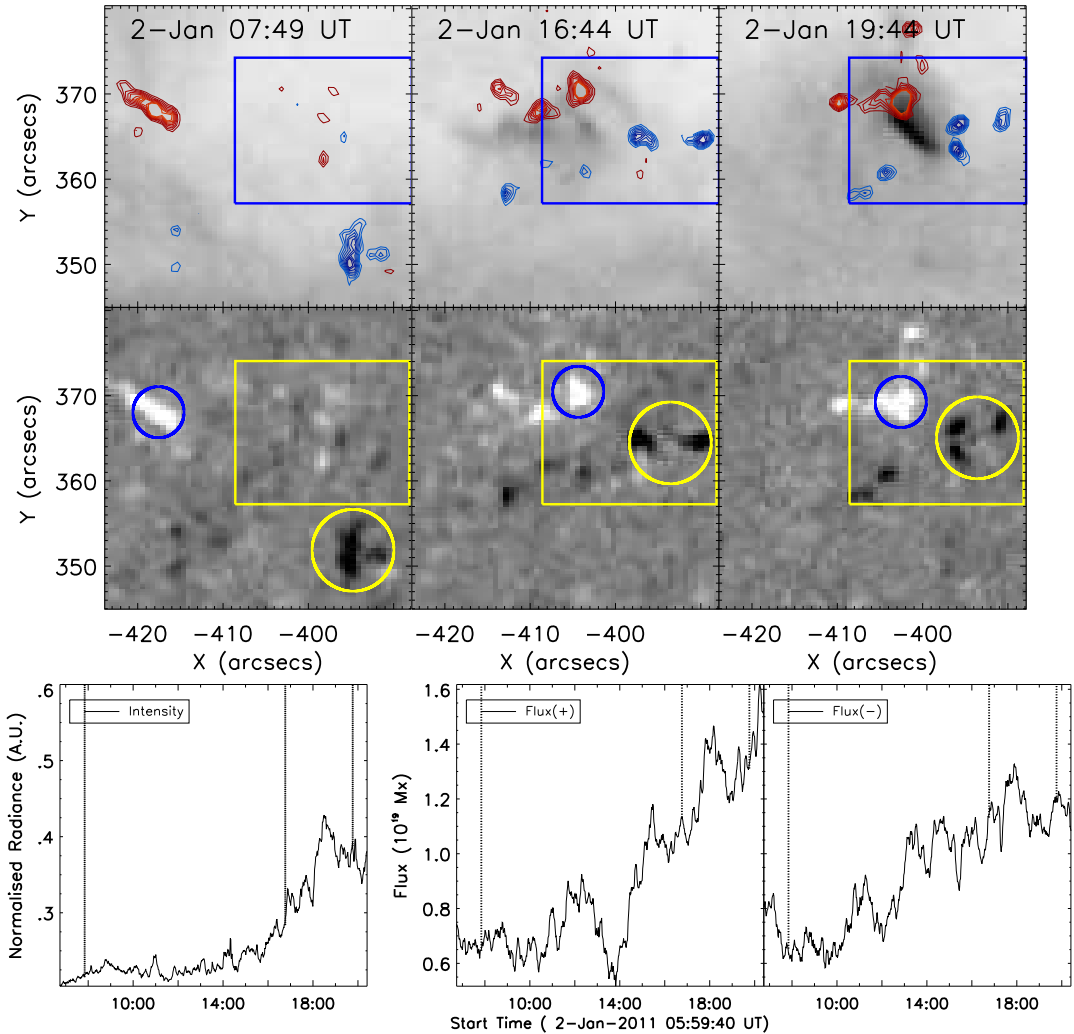}
\caption{Same as Fig.\,\ref{fig_emerg}, but for the event numbered  42 in Table\,\ref{tab_analysis}
whose associated MBF is formed through convergence. An animation is enclosed online.}
\label{fig_conv}
\end{figure*}

\subsubsection{A MBF formed by convergence}
In Fig.\,\ref{fig_conv}, we give an example of a MBF formed by convergence. The MBF is associated
with a BP seen in AIA 193\,\AA\ from 15:39\,UT on January 2 to 07:20\,UT on January 3, with a
lifetime of about 16 hours. The BP is listed as the No. 42 event in Table\,\ref{tab_analysis}.
Initially, the two polarities are found to be at a distance of about 23\arcsec\ (see image at
07:49\,UT in Fig.\,\ref{fig_conv}). Then they move towards each other. When the BP starts to be
seen in AIA 193\,\AA\ around 16:44\,UT, the distance between  the two polarities is about
6\arcsec. The distance is found to be about 4\arcsec\ when the BP reaches its emission peak in
the AIA 193\,\AA\ channel (see images at 19:44\,UT in Fig.\,\ref{fig_conv}). The 6\arcsec\
distance between the two polarities when the BP appears is consistent with what was found by
\citet{2003A&A...398..775M}.

\begin{figure*}[!ht]
\centering
\includegraphics[clip,trim=0cm .5cm 0cm 0cm,width=17cm]{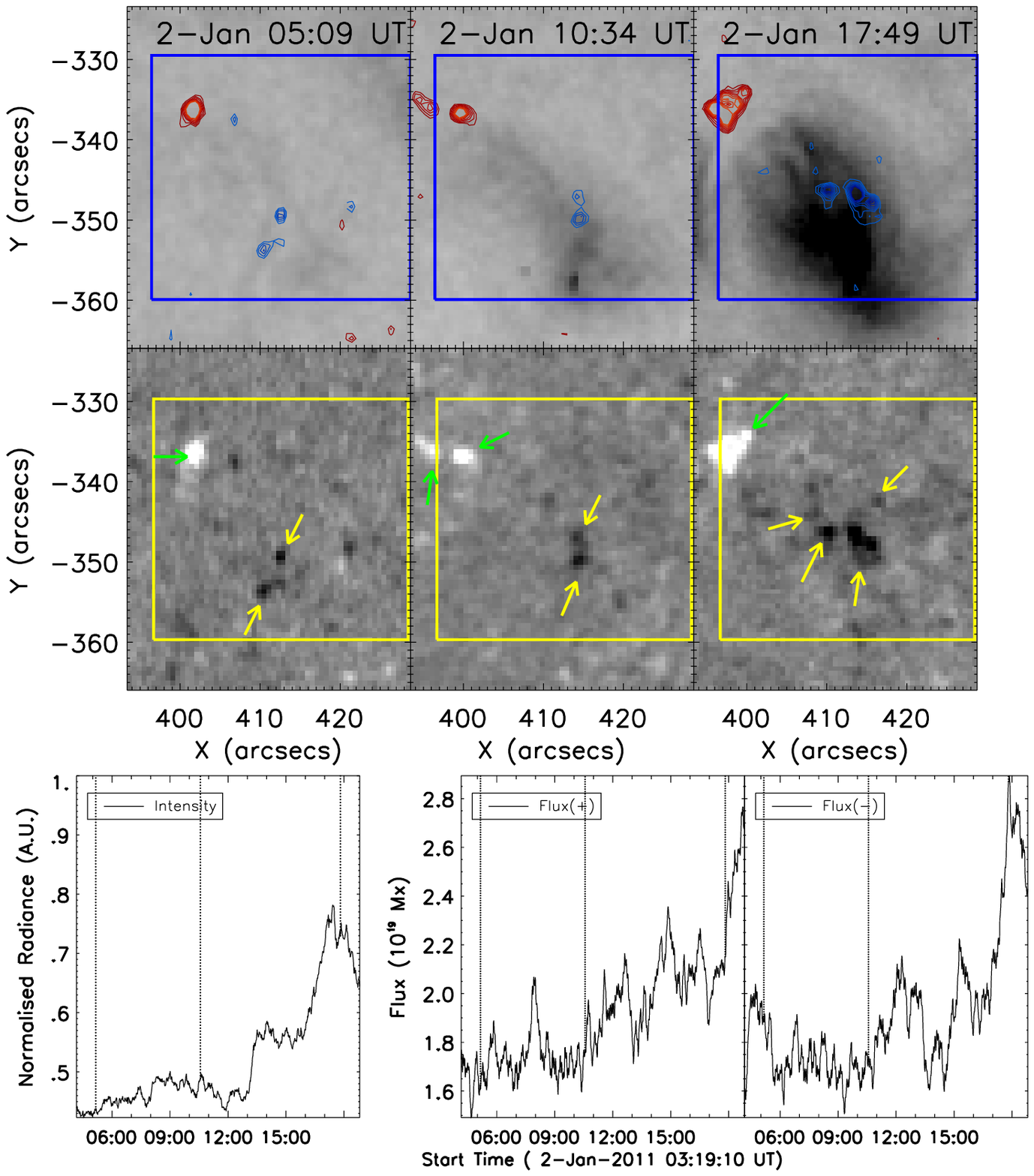}
\caption{Same as Fig.\,\ref{fig_emerg}, but for the event numbered 19 in Table\,\ref{tab_analysis}
whose associated MBF is formed through local coalescence. An animation is given online.}
\label{fig_lo_com}
\end{figure*}

\subsubsection{A MBF formed by local coalescence}
Fig.\,\ref{fig_lo_com} shows the evolution of a BP (No.~19 in Table\,\ref{tab_analysis}) and its
associated MBF formed by local coalescence. In this case a few small magnetic flux kernels are
first seen (see features denoted by arrows at 05:09\,UT in Fig.\,\ref{fig_lo_com}) surrounded by
mostly fuzzy, weak magnetic flux concentrations . The kernels coalesce
 with these  magnetic features of the same sign to form larger ones. This
process leads to the formation of a BP in  AIA\,193\,\AA. The distance between the two polarities
remains almost the same, changing slowly during the merging process. The distances measured at
05:09\,UT, 10:34\,UT and 17:49\,UT are 15\arcsec, 15\arcsec and 13\arcsec, respectively. In the
convergence cases, for instance,   the distance changes dynamically.  These weak magnetic
features can not directly trigger a BP until they merge into large ones. These cases are,
therefore, different  from the emergence cases where the flux emergence is directly followed by a
BP formation in the AIA 193\,\AA.

\subsubsection{Distance of two main polarities of a MBF}
\begin{figure*}
\centering
\includegraphics[clip,trim=0cm 0.3cm 0cm 0cm,width=17cm]{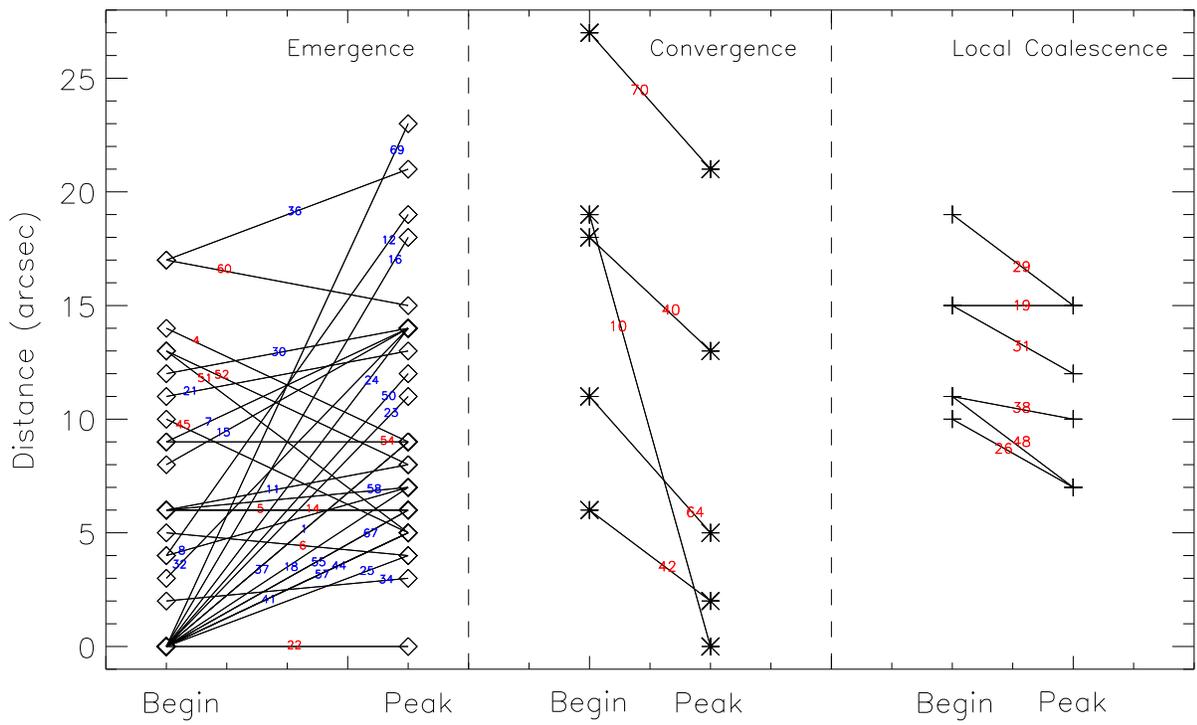}
\caption{Distances of the two main polarities of MBFs formed by emergence (left), convergence
(middle) and local coalescence at the time when the BPs are first seen (``begin'') and at
their radiation peaks (``peak'') in the AIA 193\,\AA\ passband. The denoted numbers are the numbers
of the BPs listed in Table\,\ref{tab_analysis}. The solid line connects the two distances determined
at the ``'begin'' and ``peak'' of an event.}
\label{fig_dist_var}
\end{figure*}

As already shown in the above sections, the distance of the two main polarities of the MBFs
varies in different stages of the evolution of a BP, and it seems to be dependent on the
formation mechanism of the MBFs. In Fig.\,\ref{fig_dist_var}, we show the distances at the time
when the BPs are first seen and at  their radiation peak in the AIA 193\,\AA\ channel. For MBFs
formed by convergence, the distances at the radiation peaks are shorter than that at the  time of
the first appearances. For MBFs formed by local coalescence, the distances at the radiation peaks
are also shorter than (or equal to) those at the first appearance, but the changes are small
compared to the convergence cases. The distance variations are more complex for those formed by
bipolar emergence. For the 35 bipolar emergence cases, the distances at the time of the first
appearance of  25 cases are shorter than, of  4 cases equal to, and of 6 cases  longer than those
at the radiation peaks. The variation of the distances during a BP lifetime is complex and it might randomly
increase/decrease.

\subsection{Magnetic cancellation associated with BPs}
\label{sect_mcbp} Magnetic cancellation is another phenomenon observed during a BP evolution (see
Section\,\ref{sect_intro} for reviews of previous works) and  it  has been observed in all BPs
studied here. Magnetic cancelation is a phenomenological description of a magnetic process that
interprets the disappearance and/or decrease of opposite sign magnetic features when they are in contact with
each other observed in magnetograms. As pointed out by \citet{1999SoPh..190...35H},
magnetic flux that is retracting below the solar surface in  cancellation sites  is believed to
be an outcome of magnetic reconnection\,\citep{2007mare.book.....P}. Since magnetic cancellation
is related to BPs, this suggests that a BP and its associated magnetic cancellation are likely to
be the result of magnetic reconnection. Thus, a study on how  magnetic cancellation occurs can
help to understand how
 magnetic reconnection process operates to energise a BP.

\par
In our study, we found that it is difficult to identify magnetic cancellation only by following
the temporal variation of the magnetic flux in the lightcurves of the total positive and negative
fluxes associated with a BP due to the complex  magnetic flux  evolution in the photosphere.
Because of the  non-potentiality of the magnetic field and the lack of data that are suitable for
a non-linear magnetic field extrapolation, potential field extrapolation models will not
reproduce the real field configuration of BPs.   Thus, the only reliable analysis at present
remains a visual inspection of longitudinal magnetograms.

From the analysis of  the  70 samples, we found that magnetic cancellation associated  with a BP
can occur in three different ways: (\I) between a MBF and  small weak magnetic features that are
evolving in the vicinity; (\II) within a MBF when  the two main polarities are moving towards
each other from a large distance; (\III) within a MBF  whose two main polarities emerge in the same
place simultaneously.

\par
\begin{figure*}[!ht]
\centering
\includegraphics[trim=3cm 2.5cm 2cm 2.2cm,clip,width=5.5cm]{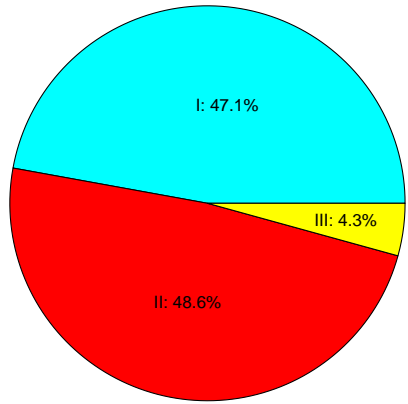}
\caption{A pie-chart to show the three different ways of evolutionary pattern that magnetic cancellation occurs along with 70 studied BPs. }
\label{fig_cal_stat}
\end{figure*}

Magnetic cancellation of category \I\ and \II\ are found for the majority of our samples.  We
found that  67 out of 70 cases occur in these two ways of evolutionary pattern. We also noticed
that most of the BPs do not follow only a single  way. Cancellation associated with categories
\I\ and \II\ were found at  different time during  the evolution of  BPs. However, for any
individual BP, we could always find one dominant pattern in the cancellation, which is found to
occur during most of the BP lifetime. This dominant evolutionary pattern  is listed in
Table\,\ref{tab_analysis}. In total, we found that 33 BPs are primarily associated with magnetic
cancellation of category \I\ and  34 BPs of category \II. Fig.\,\ref{fig_cal_stat} presents a
pie-chart of  the statistical results of three categories of cancellation.
\par
We also further investigated whether magnetic cancellation occurring in the three different ways
of evolutionary patterns affects the BP lifetimes. Because there are very few cases that are
related to category \III, we only considered the first two categories. In category \I, the
lifetimes vary from 6.4 hours to 45 hours with an average of 23 hours, and from 2.7 hours to 58.8
hours with an average of 19.6 hours in category \II. Although the average lifetime of the BP in
category \I\ is a few hours longer, there is no clear correlation between the BP lifetimes and
the magnetic cancellation categories. In our investigation, a BP will disappear once one of the
main polarities becomes too weak or disappears.

\begin{figure*}[!ht]
\centering
\includegraphics[clip,trim=0cm 1.5cm 0cm 1cm,width=17cm]{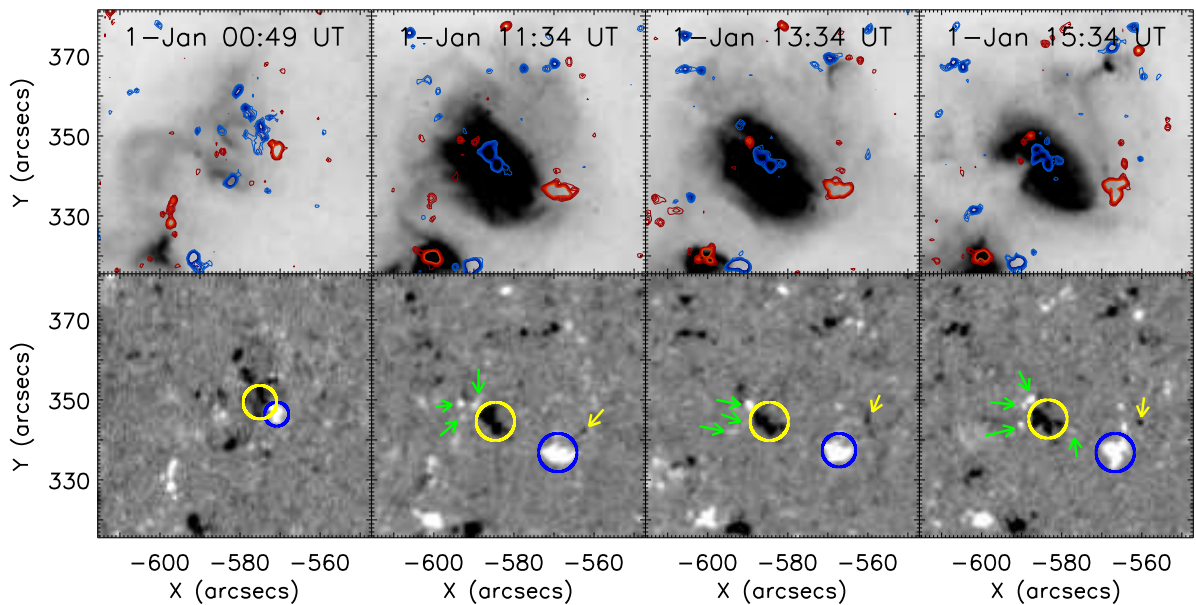}
\caption{An example (No. 1 listed in Table\,\ref{tab_analysis}) of a BP that is associated with
magnetic cancellation occurring between the main polarities of the MBF and small weak features.
The BP seen in AIA 193\,\AA\ is given in the top row, and its associated magnetograms are given in the
bottom row. The contours of positive (red) and negative (blue) magnetic flux density are over-plotted in the top row.
The blue and yellow circles mark the main MBF. The yellow and green arrows
indicate that the small magnetic features are about to cancel with the polarities of the MBF. This is the
same event as shown in Fig.\,\ref{fig_emerg} and an animation of this BP is attached to
Fig.\,\ref{fig_emerg}.}
\label{fig_small}
\end{figure*}

\begin{figure*}[!ht]
\centering
\includegraphics[clip,,trim=0cm 0.3cm 0cm 0cm, width=17cm]{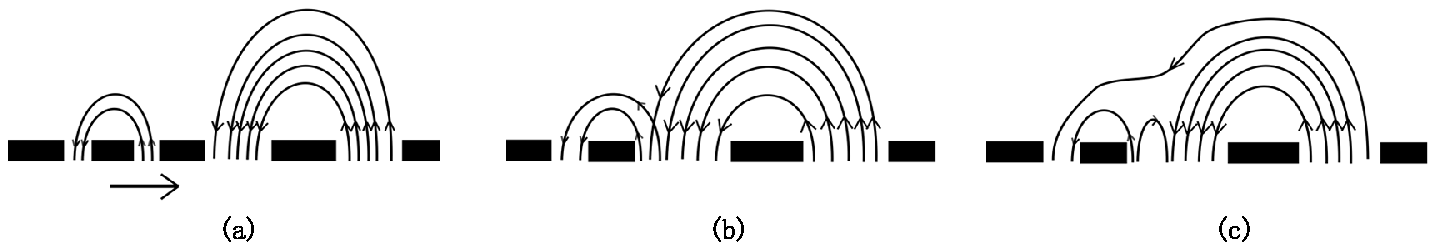}
\caption{A cartoon demonstrating the topological evolution of the magnetic fields which are involved in
the magnetic cancellation and BP heating in the case shown in Fig.\,\ref{fig_small}. Panel (a)
presents an initial magnetic topology with a large magnetic loop (corresponding to the MBF) and a
small one (corresponding to small and weak magnetic features). Panel (b) shows
that the large magnetic loops are interacting with the small ones and magnetic reconnection occurs.
Panel (c) presents the magnetic field topology after reconnection. The produced small loop will
submerge due to the magnetic tension.}
\label{fig_md_small}
\end{figure*}

\subsubsection{A case of category (I)}
Fig.\,\ref{fig_small} shows the  evolution of a BP and its associated magnetic features, which
present magnetic cancellation occurring between the main MBF and small magnetic features. This BP
is the same as the one presented in Fig.\,\ref{fig_emerg} and the attached animation there. In
this event, small-scale weak magnetic features are found to play a key role in the BP evolution.
In Fig.\,\ref{fig_small}, we show a few snapshots of the BP and its associated magnetic features.
These small magnetic features move towards and interact with the main polarities. This
 process is  followed by the disappearance of the small features and is accompanied by
emission enhancements in AIA 193\,\AA. This example suggests that  the main MBF builds up
the skeleton of the BP but the activity of the  small-scale magnetic fluxes  is responsible for
the BP heating.

\par
Based on the observations shown in Fig.\,\ref{fig_small}, we suggest a cartoon model
 showing a reconnection process that can explain the observed magnetic cancellation and the BP formation (see Fig.\,\ref{fig_md_small}). In
this model, the large loops are rooted in the main MBF  while the small loops connect the small
magnetic features
 (Fig.\,\ref{fig_md_small}a). The two loop systems move towards each
other due to the solar convection. They interact and reconnect with each other
(Fig.\,\ref{fig_md_small}b), which then releases energy to heat the loop system that represents a
BP. The reconnection results in a tiny loop that will submerge following the reconnection
(Fig.\,\ref{fig_md_small}c). In this model, the reconnection  may occur in the lower solar
atmosphere since a small loop is involved. Evidences of magnetic reconnection between two loop
systems in the solar atmosphere have been reported by numerous studies, e.g. \citet{huang2014}
and \citet{2015arXiv150707594H}.

\begin{figure*}[!ht]
\centering
\includegraphics[clip,trim=0cm 1.5cm 0cm 1cm,width=17cm]{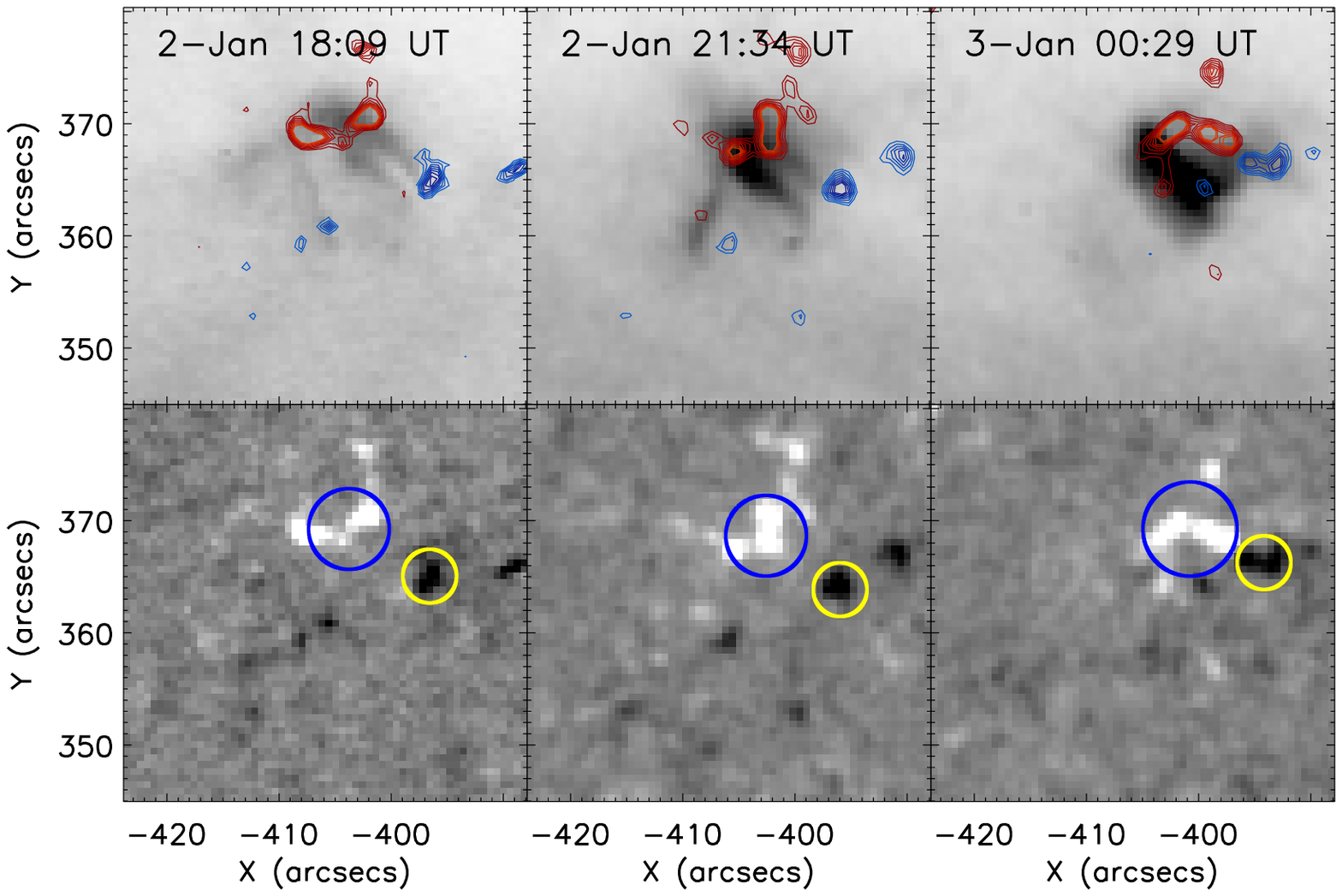}
\caption{Same as Fig.\,\ref{fig_small}, but gives an example of magnetic cancellation occurring between
two main polarities of the MBF when they are moving towards each other from a far distance.
Its associated BP is numbered as 42 in Table\,\ref{tab_analysis}.This is the same event as shown in
Fig.\,\ref{fig_conv}, and an animation is given online attached
to Fig.\,\ref{fig_conv}.}
\label{fig_converge}
\end{figure*}

\begin{figure*}[!ht]
\centering
\includegraphics[clip,trim=0cm 0.3cm 0cm 0cm,width=17cm]{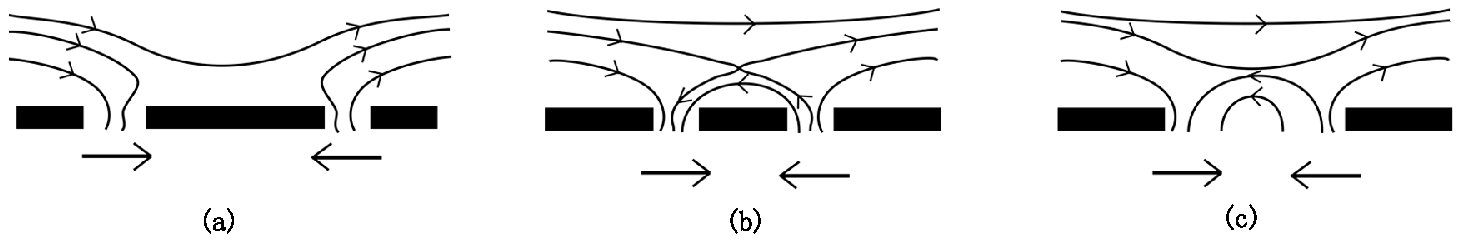}
\caption{A cartoon presenting the evolution of the magnetic field that can explain the magnetic
cancellations and the BP shown in Fig.\,\ref{fig_converge}. Panel (a) shows that two
polarities in the MBF belong to two different loop systems. The convergence motion forces them
to move closer, which leads to magnetic reconnection shown in panel (b). The magnetic
reconnection heats the BP and the generating small loop will submerge (panel c).}
\label{fig_md_converge}
\end{figure*}

\subsubsection{A case of category (II)}
Fig.\,\ref{fig_converge} shows an example where  a BP is flared up during  the converging motions
of the two main polarities. This BP is the same as the one shown in Fig.\,\ref{fig_conv} and its
full evolution is given in the attached animation. The event clearly demonstrates that the
emission in AIA 193\,\AA\ increases when the two main polarities move closer and while are
 in contact (see image at 00:29\,UT in Fig.\,\ref{fig_converge}). The magnetic
cancellation is found to occur between the two main polarities. When the fluxes of  the main
polarities are canceled out, the BP disappears. A similar relation between magnetic  flux
cancellation and BP evolution  has been reported in  several
studies\,\citep[e.g.][etc.]{2001SoPh..201..305B,2003A&A...398..775M,2012A&A...548A..62H}.

\par
 Although two cancelling polarities might connect each other in the initial stage as pointed out
in Section\,\ref{sect_mbf_form},
 we assume here that they do not, and then this case can be explained by a BP scenario proposed by \citet{1994ApJ...427..459P}.  We give a simplified version of their model in
Fig.\,\ref{fig_md_converge}. The initial configuration of the model  consists of two magnetic
loop systems. Each of them has one footpoint rooted in one of the polarities of the main MBF
(Fig.\,\ref{fig_md_converge}a). The converging motion forces the loops to move closer and their
field lines to reconnect (Fig.\,\ref{fig_md_converge}b). The resulting small magnetic loops
connecting the two main polarities of MBF then submerge. The fluxes in the main polarities are
canceled out by this process (Fig.\,\ref{fig_md_converge}c). In this scenario, a BP is produced
in the reconnection site and the resulting long loop is not  entirely heated by the reconnection
process. For more details on the convergence model (Fig.\,\ref{fig_md_converge}) please refer to
\citet{1994ApJ...427..459P} and their follow-up
works\,\citep{1994SoPh..153..217P,2006MNRAS.366..125V,2006MNRAS.369...43V}.

\begin{figure*}[!ht]
\centering
\includegraphics[clip,trim=0cm 1cm 0cm 0.5cm,width=17cm]{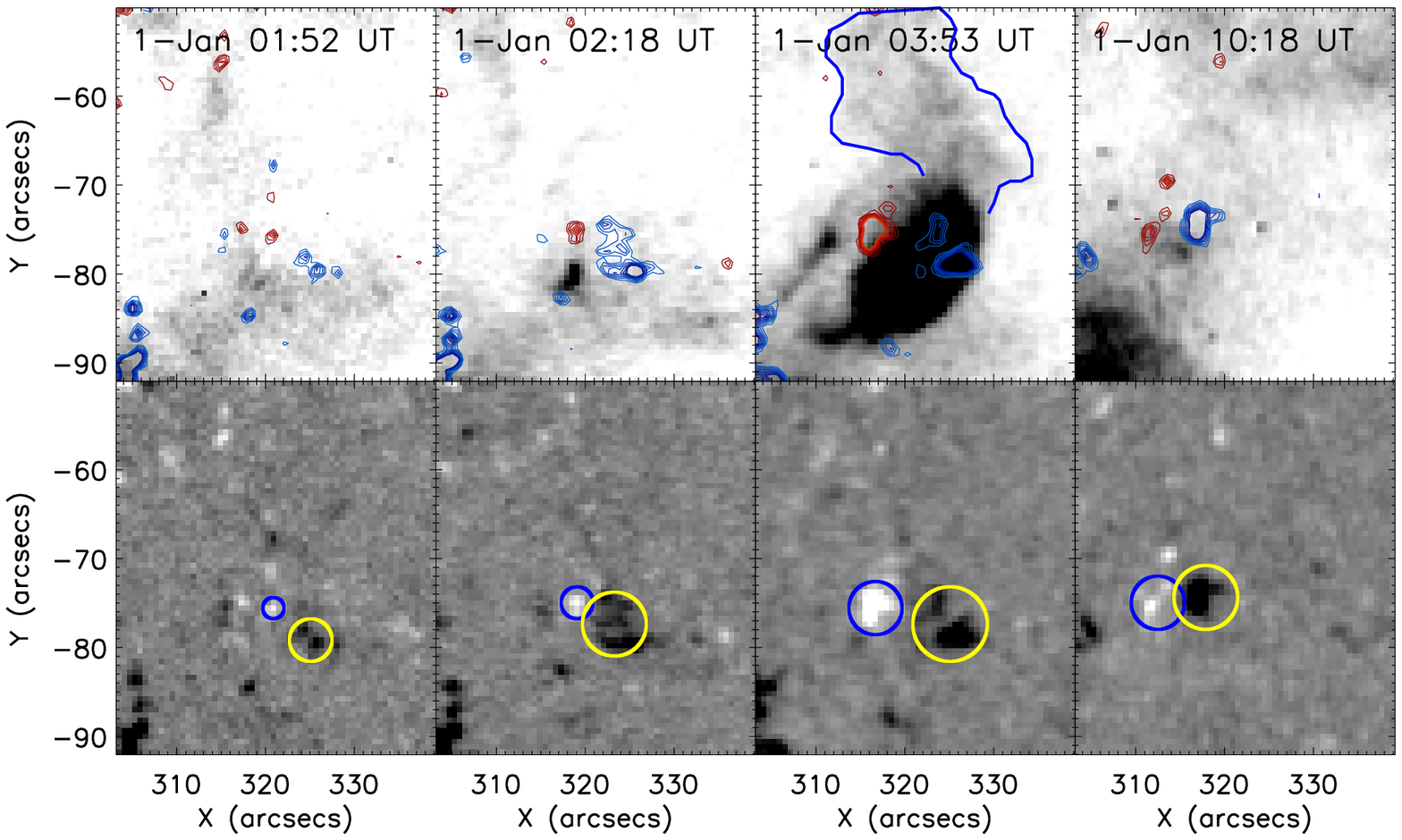}
\caption{An example showing the evolution of a BP (No. 22 listed in Table\,\ref{tab_analysis}) and its
associated magnetic features. This is an example whose associated magnetic cancellation occurs within two main polarities of the MBF that emerged from the same location simultaneously. The BP images seen in AIA 193\,\AA\ are given in the top row and
the magnetograms are given in the bottom row. The blue and yellow circles mark the polarities of
the MBF. The contours of positive (red) and negative (blue) magnetic flux density are over-plotted in the top row. The blue line in the AIA image at 03:53\,UT denotes a blowout plasma structure originated from the BP.
An animation is given online.}
\label{fig_CME}
\end{figure*}

\par
\begin{figure*}
\centering
\includegraphics[clip,trim=0cm 0.3cm 0cm 0cm,width=17cm]{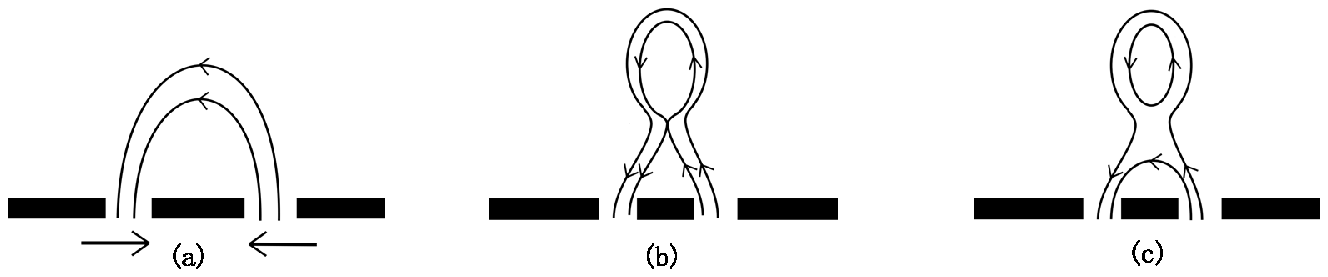}
\caption{A cartoon showing a magnetic reconnection process that may be responsible for the magnetic
cancellation and the plasma ejection presented in Fig.\,\ref{fig_CME}. Panels (a), (b) and
(c) display the initial stage, the reconnection and the resultant magnetic field.}
\label{fig_md_CME}
\end{figure*}

\subsubsection{A case of category (III)}
Magnetic cancellation occurring in category \III\ is  relatively rare (only 3 out of 70 cases).
In this category, the magnetic cancellation happens between the two main polarities of a MBF.
However, it differs from the category \II\ because here the two main polarities emerge in the
same location simultaneously (i.e. bipolar emergence). This indicates that magnetic loops have
been connecting them since their birth. In this  case, the emerged main polarities first move
away from each other and  a BP appears in the images of the AIA 193\,\AA\ channel.  Further in
the evolution of the BP the two main polarities  move towards each other which initiates
magnetic cancellation.

\par
Fig.\,\ref{fig_CME} displays a BP (No.~22 in Table\,\ref{tab_analysis}) which belongs to category
\III. The BP starts to be seen in AIA 193\,\AA\ at 02:18\,UT on January 1 while its  MBF is
emerging. The two polarities of the MBF are  at zero distance, i.e. the two polarities touch each
other.  While the two main polarities move away from each other, the BP increases in both size
and emission. At 03:20\,UT, the two main polarities have a maximum separation of about 10\arcsec.
After, they start to move towards each other, and magnetic cancellation occurs between them,
while the BP  becomes  brighter. We also observed a shearing motion that occurs while  the two
polarities move  closer  (see Fig.\,\ref{fig_CME}). The BP reaches its emission peak at around
03:50\,UT when the two main polarities are again in contact. The BP lasts until 10:18\,UT when
the flux in the positive polarity is nearly canceled out.

\par
To determine whether the magnetic cancellation observed  in this case is associated with magnetic
reconnection  that energises  the BP can be challenging. In Fig.\,\ref{fig_md_CME}, we present a
cartoon layout of magnetic fields involved in magnetic reconnection that can generate energy to
heat the BP. In this cartoon, the magnetic loops are initially connecting the two main polarities
(Fig.\,\ref{fig_md_CME}a). Converging motions then force the field lines to reconnect
(Fig.\,\ref{fig_md_CME}b) due to  instabilities (for example, kink instability due to shearing
motion). The reconnection will result in a smaller loop that will submerge and a magnetic island
that will be ejected from the reconnection site (Fig.\,\ref{fig_md_CME}c). This model can be
considered as a micro version of a flare model proposed by \citet{1974SoPh...34..323H}. The
ejected magnetic island is the main phenomenon that can be used to test the scenario. It can
generate a so-called blowout jet\,\citep[see example given in][]{2014SoPh..289.3313Y} or a
mini-CME\,\citep[see example given in][]{2010A&A...517L...7I}. We then searched for obseravtional
signatures of such a phenomenon by tracing the evolution of the BP in AIA 193\,\AA. Ejections of
plasma similar to blowout jet are present from time to time throughout the BP lifetime duration
(see online animation attached to Fig.\,\ref{fig_CME}). An example is given in the image at
03:53\,UT of Fig.\,\ref{fig_CME}, where the front of a fuzzy arch-like feature is outlined. The
structure is moving away from the BP, and that can be followed in the attached online animation.
Observational evidence for instabilities are needed to fully prove this scenario, but this  is
beyond the scope of the current study.

\section{Conclusion and Summary}
\label{sect_conc}

In the present study,  we aim to investigate how a BP-associated MBF is formed and how magnetic
cancellation associated  with BPs occurs. We have traced the evolution of 70 BPs throughout their
lifetimes together with their associated photospheric magnetic flux  using SDO/AIA and SDO/HMI
observations.

\par
We found that the lifetimes of the BPs vary from 2.7 to 58.8 hours averaging at 20.9 hours. The
distance of the two opposite polarities of a MBF associated with these BPs changes throughout the
BP lifetime. When the BPs first appear, the distance distributes in the range from  0 to
31.5\arcsec. A zero distance is found in 16 BP-associated MBFs from which 15 are related to
bipolar flux emergence. The distance spreads from 0\arcsec to 26.8\arcsec with a a  peak of the
distribution at 5\arcsec when the BPs are at their radiation peak.

\par
We found that a BP-associated MBF can be formed in three ways: emergence where a magnetic feature
appears and grows as seen in the HMI magnetograms, convergence where the two pre-existing
polarities move towards each other from a large distance, and local coalescence where a magnetic
feature is formed by mergence  with several small magnetic flux concentrations  of the same sign.

Out of the 70 cases, flux emergence is the main process of a MBF buildup for 52 BPs, mainly
convergence is seen in 28 BPs, and 14 cases are associated with local coalescence.
 For those BPs whose two polarities are purely formed by the same mechanism  (46 cases),
  35 cases are related to bipolar emergence, 5 cases are formed by
convergence and 6 cases by local coalescence. In the 35 cases, the time difference between
first appearance of the  BPs in AIA 193\,\AA\ and the flux emergence ranges from 0.1 to 3.2 hours
with an average of 1.3 hours.

We further investigated the distance between  the  two main polarities of a MBF at the time when
the BPs were first visible and when they were  at their radiation peak in the AIA 193\,\AA\
channel. In the cases of convergence we found that the distance at the time when the BP is first
visible is much longer than that at the radiation peak. In the cases of local coalescence, the
distances are  more or less the same. In  a total of 35 cases of bipolar emergence, the distances
at the first appearance of 25 are shorter than, 4 are equal to, and 6 are longer than that at the
BP's radiation peaks. The variation of the distances during a BP lifetime might randomly
increase/decrease.

\par
All 70 BPs are found to be associated with magnetic cancellation. However, magnetic cancellation
associated with  BPs can  occur in three ways: (\I) between the main polarities of a MBF and
small weak magnetic fields, (\II) between the two polarities of a main MBF while they are moving
towards each other from a large distance, and (III) between two polarities of a MBF that emerge
together at  the same location. Out of 70 BPs, in 33 cases magnetic cancellation primarily occurs
in category \I, in 34 cases in category \II,  and 3 cases belong to category \III. For each
category, we displayed a cartoon model that can explain the observed magnetic cancellation and
reconnection responsible for  the heating of the BPs. While the main polarities of a
BP-associated MBF build up the skeleton of a BP, we find that the magnetic activities responsible
for the BP heating may involve small weak fields.

\acknowledgments {\it Acknowledgments:} We thank very much the anonymous referee for his/her
helpful and constructive comments. We also thank Dr. Armin Theissen for his linguistic
corrections. We thank Dr. Kamalam Vanninathan and Dr. Kalugodu Chandrashekhar for carefully
reading the manuscript, and their useful comments and suggestions. This research is supported by
the China 973 program 2012CB825601,  the National Natural Science Foundation of China under
contracts: 41404135 (ZH), 41274178 and 41474150 (LX \& ZH), and 41174154, 41274176 and 41474149
(BL), the Shandong provincial Natural Science Foundation ZR2014DQ006, China Postdoctoral Science
Foundation funded project and special funds from Shandong provincial postdoc innovation project
(ZH). MM is supported by the Leverhulme Trust. Research at the Armagh Observatory is grant-aided
by the N. Ireland Department of Culture, Arts and Leisure. AIA and HMI data is courtesy of SDO
(NASA). We thank JSOC for providing downlinks of the SDO data.

{\it Facilities:} \facility{SDO}.

\bibliographystyle{apj}
\bibliography{bibliography}

\newpage
\begin{onecolumn}
\begin{scriptsize}
\begin{longtable}{|c|c|c|c|c|c|c|c|c|c|}
\caption{The studied BPs, formations of their associated MBFs, and magnetic cancellations
associated with heating.}
\label{tab_analysis}\\
\hline
&Location&&&&\multicolumn{2}{c|}{Bipolar}&&\multicolumn{2}{c|}{Distance of main polarities}\\
No. &(x,y)& Start Time & End time &Duration&\multicolumn{2}{c|}{formation\tablenotemark{\spadesuit}}&Cancel-&\multicolumn{2}{c|}{(arcsec)}\\
\cline{6-7} \cline{9-10}
&[arcsec]&&&(hours)& Positive & Negative &lation\tablenotemark{\clubsuit}& Beginning\tablenotemark{\ast}& Peak\tablenotemark{\star}\\
\hline
\endfirsthead

\multicolumn{10}{c} {{\bfseries \tablename\ \thetable{} -- continued}}\\
\hline
&Location&&&&\multicolumn{2}{c|}{Bipolar}&&\multicolumn{2}{c|}{Distance of main polarities}\\
No. &(x,y)& Start Time & End time &Duration&\multicolumn{2}{c|}{formation\tablenotemark{\spadesuit}}&Cancel-&\multicolumn{2}{c|}{(arcsec)}\\
\cline{6-7} \cline{9-10}
&[arcsec]&&&(hours)& Positive & Negative &lation\tablenotemark{\clubsuit}& Beginning\tablenotemark{\ast}& Peak\tablenotemark{\star}\\
\hline
\endhead

\hline \multicolumn{10}{r}{{Continued on next page}}
\endfoot

\hline \hline
\endlastfoot
1 & $-$542, 338 & 01-01 00:04 & 01-02 13:19 & 37.3 & A & A & \I & 0 & 9\\
\hline
2 & $-$289, 490 & 01-01 22:26 & 01-03 23:58 & 49.5 & C+B & B & \II & 32 & 27\\
\hline
3 & 310, $-$119 & 01-01 10:29 & 01-03 21:19 & 58.8 & B & A+B & \II & 12 & 9\\
\hline
4 & 130, $-$169 & 01-02 18:09 & 01-03 09:00 & 14.9 & A & A & \I & 14 & 9\\
\hline
5 & 60, $-$134 & 01-01 07:03 & 01-02 00:10 & 17.1 & A & A & \I & 6 & 6\\
\hline
6 & $-$34, $-$204 & 01-01 11:53 & 01-02 23:50 & 36.0 & A & A & \I & 5 & 4\\
\hline
7 & 170, $-$229 & 01-01 14:43 & 01-02 18:50 & 28.1 & A & A & \I & 9 & 14\\
\hline
8 & $-$109, $-$209 & 01-01 20:00 & 01-02 16:40 & 20.7 & A & A & \I & 4 & 7\\
\hline
9 & 140, $-$109 & 01-02 21:39 & 01-03 23:59 & 26.3 & A+B & B & \I & 6 & 8\\
\hline
10 & $-$89, $-$69 & 01-01 14:58 & 01-02 07:10 & 16.2 & B & B & \II & 19 & 0\\
\hline
11 & 95, $-$164 & 01-01 11:24 & 01-02 12:30 & 25.1 & A & A & \I & 6 & 8\\
\hline
12 & 205, $-$94 & 01-02 04:04 & 01-03 05:40 & 25.6 & A & A & \I & 4 & 19\\
\hline
13 & $-$59, $-$449 & 12-31 23:48 & 01-01 17:30 & 17.7 & B & A+B & \II & 8 & 8\\
\hline
14 & 340, $-$389 & 01-01 16:25 & 01-02 11:30 & 19.1 & A & A & \III & 6 & 6\\
\hline
15 & 180, $-$224 & 01-01 14:58 & 01-02 20:10 & 29.2 & A & A & \I & 8 & 14\\
\hline
16 & 155, $-$329 & 01-01 20:15 & 01-03 08:10 & 35.9 & A & A & \I & 0 & 18\\
\hline
17 & 255, $-$369 & 01-02 11:54 & 01-03 01:20 & 13.4 & B & A+B & \II & 8 & 3\\
\hline
18 & 275, $-$319 & 01-02 06:44 & 01-03 09:10 & 26.4 & A & A & \I & 0 & 7\\
\hline
19 & 480, $-$339 & 01-02 12:54 & 01-03 09:00 & 20.1 & C & C & \II & 15 & 15\\
\hline
20 & 270, $-$359 & 01-02 10:14 & 01-03 03:50 & 17.6 & B & A+B & \II & 13 & 19\\
\hline
21 & 290, $-$99 & 12-31 22:10 & 01-01 23:30 & 25.3 & A & A & \I & 11 & 13\\
\hline
22 & 340, $-$79 & 01-01 02:12 & 01-01 10:19 & 8.1 & A & A & \III & 0 & 0\\
\hline
23 & 660, $-$169 & 01-02 11:19 & 01-03 06:40 & 19.4 & A & A & \I & 0 & 11\\
\hline
24 & 630, $-$179 & 01-02 17:09 & 01-03 07:30 & 14.4 & A & A & \I & 0 & 14\\
\hline
25 & 770, 45 & 01-02 17:59 & 01-03 12:00 & 18.0 & A & A & \I & 0 & 4\\
\hline
26 & 160, $-$144 & 01-01 03:08 & 01-02 00:50 & 21.7 & C & C & \II & 10 & 7\\
\hline
27 & 440, $-$164 & 01-01 11:18 & 01-02 00:20 & 13.0 & C+B & B & \II & 9 & 4\\
\hline
28 & 740, 0 & 01-02 15:28 & 01-03 04:00 & 12.5 & C+B & B & \II & 13 & 5\\
\hline
29 & 520, 30 & 01-01 23:10 & 01-02 19:50 & 20.7 & C & C & \II & 19 & 15\\
\hline
30 & 510, $-$49 & 01-02 19:34 & 01-03 17:00 & 21.4 & A & A & \I & 12 & 14\\
\hline
31 & 510, $-$129 & 01-02 21:35 & 01-03 09:20 & 11.8 & C & C & \II & 15 & 12\\
\hline
32 & 645, $-$64 & 01-03 00:59 & 01-03 15:30 & 14.5 & A & A & \I & 3 & 14\\
\hline
33 & 100, $-$169 & 12-31 21:58 & 01-01 20:40 & 22.7 & B & A+B & \II & 16 & 7\\
\hline
34 & $-$359, 510 & 01-01 07:48 & 01-01 14:29 & 6.7 & A & A & \III & 2 & 3\\
\hline
35 & $-$264, 345 & 01-01 16:00 & 01-03 01:40 & 33.7 & A+B & B & \II & 20 & 9\\
\hline
36 & $-$234, 380 & 01-02 00:25 & 01-03 06:20 & 29.9 & A & A & \I & 17 & 21\\
\hline
37 & $-$199, 380 & 01-01 17:35 & 01-02 11:20 & 17.8 & A & A & \I & 0 & 9\\
\hline
38 & $-$509, 475 & 01-01 08:48 & 01-01 21:50 & 13.0 & C & C & \II & 11 & 10\\
\hline
39 & $-$339, 400 & 01-01 00:43 & 01-03 03:30 & 50.8 & B & A+B & \II & 0 & 12\\
\hline
40 & $-$124, 405 & 01-03 08:24 & 01-03 23:59 & 15.6 & B & B & \II & 18 & 13\\
\hline
41 & $-$359, 335 & 01-02 10:49 & 01-03 17:00 & 30.2 & A & A & \I & 0 & 5\\
\hline
42 & $-$319, 370 & 01-02 15:39 & 01-03 07:20 & 15.7 & B & B & \II & 6 & 2\\
\hline
43 & $-$629, 340 & 01-01 03:33 & 01-01 17:10 & 13.6 & B & C+B & \II & 14 & 9\\
\hline
44 & $-$469, 540 & 01-02 00:30 & 01-02 19:40 & 19.2 & A & A & \I & 0 & 5\\
\hline
45 & $-$549, 320 & 12-31 17:58 & 01-01 23:20 & 29.4 & A & A & \I & 10 & 5\\
\hline
46 & $-$149, 430 & 01-02 16:09 & 01-03 02:00 & 9.9 & A+B & B & \II & 11 & 4\\
\hline
47 & 0, 580 & 01-03 03:59 & 01-03 12:50 & 8.9 & C+B & B & \II & 18 & 11\\
\hline
48 & $-$454, 145 & 01-01 09:48 & 01-01 18:50 & 9.0 & C & C & \II & 11 & 7\\
\hline
49 & $-$484, 125 & 01-01 08:38 & 01-02 01:50 & 17.2 & A+B & B & \II & 4 & 9\\
\hline
50 & $-$379, 205 & 12-31 20:29 & 01-01 20:20 & 23.9 & A & A & \I & 0 & 12\\
\hline
51 & $-$344, 295 & 01-01 00:38 & 01-01 17:40 & 17.0 & A & A & \I & 13 & 5\\
\hline
52 & $-$449, 235 & 12-31 13:59 & 01-02 17:30 & 51.5 & A & A & \I & 13 & 8\\
\hline
53 & $-$429, 205 & 01-01 14:43 & 01-02 01:40 & 11.0 & A+B & B & \II & 16 & 0\\
\hline
54 & $-$349, 180 & 01-01 23:04 & 01-02 14:50 & 15.8 & A & A & \I & 9 & 9\\
\hline
55 & $-$289, 90 & 01-02 07:35 & 01-02 14:00 & 6.4 & A & A & \I & 0 & 6\\
\hline
56 & $-$194, 210 & 01-02 13:39 & 01-02 17:40 & 4.0 & A+B & B & \II & 6 & 5\\
\hline
57 & $-$239, 310 & 01-02 12:04 & 01-03 14:50 & 26.8 & A & A & \I & 0 & 5\\
\hline
58 & $-$174, 165 & 01-02 12:18 & 01-03 01:30 & 13.2 & A & A & \I & 6 & 7\\
\hline
59 & $-$14, 305 & 01-03 02:09 & 01-03 20:40 & 18.5 & B & A+B & \II & 16 & 12\\
\hline
60 & $-$79, 260 & 01-02 18:08 & 01-03 18:00 & 23.9 & A & A & \I & 17 & 15\\
\hline
61 & $-$254, 315 & 01-01 22:55 & 01-02 23:30 & 24.6 & B & A+B & \II & 16 & 6\\
\hline
62 & $-$479, 210 & 01-01 10:18 & 01-01 12:59 & 2.7 & A+B & B & \II & 12 & 5\\
\hline
63 & $-$434, 295 & 12-31 15:49 & 01-01 12:09 & 20.3 & C+B & B & \II & 22 & 6\\
\hline
64 & $-$469, 240 & 01-01 06:28 & 01-01 15:50 & 9.4 & B & B & \II & 11 & 5\\
\hline
65 & $-$249, 230 & 01-01 09:08 & 01-01 23:20 & 14.2 & B & A+B & \II & 14 & 11\\
\hline
66 & $-$49, 800 & 01-01 06:03 & 01-02 10:40 & 28.6 & C & A & \II & 11 & 3\\
\hline
67 & $-$181, 655 & 01-01 16:30 & 01-02 14:10 & 21.7 & A & A & \I & 0 & 6\\
\hline
68 & 45, 705 & 01-02 04:24 & 01-03 04:10 & 23.8 & C+B & B & \II & 28 & 24\\
\hline
69 & 20, 690 & 01-02 18:04 & 01-03 05:30 & 11.4 & A & A & \I & 0 & 23\\
\hline
70 & $-$144, 615 & 01-01 21:15 & 01-02 14:20 & 17.1 & B & B & \II & 27 & 21\\
\hline
\multicolumn{10}{|c|}{$\spadesuit$ Formation of a MBF. A: emergence; B: convergence; C: local coalescence}\\
\multicolumn{10}{|c|}{$\clubsuit$ Magnetic cancellation associated with BP heating occurs:}\\
\multicolumn{10}{|c|}{\I: between the MBF and small weak fields;}\\
\multicolumn{10}{|c|}{ \II: within the main polarities of a MBF moving from far distance;}\\
\multicolumn{10}{|c|}{ \III: within the main polarities of a MBF emerging in the same location.}\\
\multicolumn{10}{|c|}{$\ast$ Beginning: when a BP starts to be seen in AIA\,193\,\AA.}\\
\multicolumn{10}{|c|}{$\star$ Peak: when a BP is seen at its emission peak in AIA\,193\,\AA.}
\end{longtable}
\end{scriptsize}
\end{onecolumn}

\end{document}